\documentclass{aastex631}

\usepackage{float} 

\shorttitle{Pietras et al. 2023}
\shortauthors{Pietras et al.}
\graphicspath{{./}{figures/}}

\begin{document}

\title{Analysis of Solar-like X-Class Flare on Wolf 359 Observed Simultaneously with TESS and XMM-Newton}

\author[0000-0002-8581-9386]{M. Pietras}
\affiliation{Astronomical Institute, University of Wrocław, Kopernika 11, 51-622 Wrocław, Poland}

\author[0000-0003-1853-2809]{R. Falewicz}
\affiliation{Astronomical Institute, University of Wrocław, Kopernika 11, 51-622 Wrocław, Poland}
\affiliation{University of Wrocław, Centre of Scientific Excellence - Solar and Stellar Activity, Kopernika 11, 51-622 Wrocław, Poland}

\author[0000-0002-5006-5238]{M. Siarkowski}
\affiliation{Space Research Centre, Polish Academy of Sciences (CBK PAN), Bartycka 18A, 00-716 Warsaw, Poland}

\author[0000-0002-5299-5404]{A. Kepa}
\affiliation{Space Research Centre, Polish Academy of Sciences (CBK PAN), Bartycka 18A, 00-716 Warsaw, Poland}

\author[0000-0003-1419-2835]{K. Bicz}
\affiliation{Astronomical Institute, University of Wrocław, Kopernika 11, 51-622 Wrocław, Poland}
\affiliation{University of Wrocław, Centre of Scientific Excellence - Solar and Stellar Activity, Kopernika 11, 51-622 Wrocław, Poland}

\author[0000-0001-8474-7694]{P. Pre\'s}
\affiliation{Astronomical Institute, University of Wrocław, Kopernika 11, 51-622 Wrocław, Poland}



\begin{abstract}

We present an analysis of a flare on the Wolf 359 star based on simultaneous observations of TESS and XMM-Newton. A stellar flare with energy comparable to an X-class solar flare is analyzed on this star for the first time. The main goal of the study was to determine whether the same physical processes drive and occur in stellar flares as in the solar flares. We tried to estimate the flare class by various direct and indirect methods. Light curves and spectra in different energy ranges were used to determine the parameters and profiles of the flare. From the XMM-Newton EPIC-pn X-ray data, we estimated the temperature and emission measure during the flare. The thermodynamical timescale and the loop semi-length were also determined with two different methods. The RGS spectra enabled us to calculate the differential emission measure (DEM) distributions. The obtained DEM distributions have three components at temperature values of 3 MK, 7 MK, and 16-17 MK. The analysis of the line ratio in helium-like triplets allowed us to determine the plasma electron density. Our results for the flare loop on Wolf 359 were compared to typical parameters for solar flares observed with GOES and RHESSI. This supports our conclusion that the processes taking place in stellar flares are like those in solar flares. The determined geometrical parameters of the phenomenon do not differ from the values of analogs occurring on the Sun.

\end{abstract}



\section{Introduction} \label{sec:intro}

Wolf 359 (CN Leo, TIC 365006789, GJ 406) is a main-sequence, single star of spectral type M6 \citep{2018MNRAS.475.1960F, 2019AJ....157...63K}. 
It is the fifth closest star to the Sun (2.4086 $\pm$ 0.003 pc \citep{2020yCat.1350....0G}). The mass and radius of this star are 0.11 \(M_\odot\) and 0.14 \(R_\odot\), respectively \citep{2021A&A...645A.100S}. Its effective temperature is about 2900 K \citep{2014ApJ...791...54G, 2018A&A...620A.180R}, although a more recent report calculated it at 2700 K \citep{2019ApJ...878..134K}. Its $\log g$ values range from 4.5 \citep{2019ApJ...878..134K} to 5.5 \citep{2013A&A...556A..15R,2018A&A...620A.180R}.  Wolf 359 is a very active, variable star \citep{2000ApJ...541..396A,2021AJ....162...11L, Lin_2022} with strong H$\alpha$ emission \citep{2006A&A...447..709P, 2015ApJ...804...64M}. It is the source of UV and X-ray emissions \citep{2007A&A...468..221F}. Modulation of its light curve is probably connected with starspots \citep{2021AJ....162...11L}. The star's rotation period is 2.7 days \citep{2019A&A...621A.126D, Guinan_2018}. The age of Wolf 359 is not well determined \citep{Guinan_2018}. The presence of planets orbiting this star is still under investigation \citep{2019arXiv190604644T, 2021AJ....162...11L, 2023A&A...670A.139R}.

Even weak stellar flares can be detected on the late spectral type stars. Due to low surface temperatures, flares significantly increase the total stellar luminosity. Stars from spectral type M3V, for masses less than 0.4 M$_\odot$, are fully convective \citep{2020ApJ...891..128M}. Their observations have provided evidence for the presence of an efficient mechanism that regenerates and amplifies magnetic fields. Examination of the alterations in magnetic activity occurring around spectral types  $\sim$M2 and $\sim$M9, yield insights into their magnetic behavior. The generation of magnetic fields persists even in fully convective objects such as very-low-mass stars and brown dwarfs. This raises questions regarding the generation of large-scale magnetic fields within fully convective objects and the potential existence of a distinct dynamo mechanism for dwarfs less massive than 0.35 M$_\odot$. This dynamo mechanism differs from the rotation-driven dynamo hypothesized to be operational in the Sun and early M-dwarfs possessing a radiative core. The decline in magnetic activity observed in late-M dwarfs, coupled with the presence of strong flares, corroborates the notion of large-scale magnetic field generation within the interior of these objects \citep{2006A&A...446.1027C}.


A stellar flare with energy comparable to a solar X-class flare is analyzed on Wolf 359 for the first time. It was possible thanks to the unprecedented sensitivity of the XMM-Newton satellite (UV and X-ray range) and TESS (white-light range). A fundamental research problem is to determine whether the same physical processes power, drive, and occur in stellar flares as in solar flares and whether the phenomena on stars run in similar scales and ways. The energies of stellar flares are much greater than the energies released by flares on the Sun. This leads to the hypothesis that either there is the scalability of stellar flares similar to the processes taking place on the Sun, or it is necessary to build a modified model of the structure, mechanisms, course, etc. of the stellar phenomena manifested by flare-like emission signatures. Observation of the solar-like X-class flare on Wolf 359 is a very crucial element on the basis of which the parameters and energy of the flare on another star with energies comparable to medium-strong solar flares were determined. Being able to observe and analyze such events is an important test of the working hypothesis that we are dealing with similar versions of the same physical phenomenon.

We present an analysis of the flare on the Wolf 359 observed on December 21, 2021. Section \ref{sec:obs} describes the sources of observation data, the TESS satellite, and the XMM-Newton telescope. The flare's light curve analysis is detailed in Section \ref{sec:analysis}. Section \ref{sec:res} contains the results of the research, the profiles of the flare, and its parameters. In Section \ref{sec:discussions} there is a discussion that includes a comparison with solar flares. Appendix \ref{sec:App1} describes the determination of the X-ray solar-like class of the flare. Appendix \ref{sec:App2} is the estimation and the uncertainties of flare loop parameters.

\section{Observations} \label{sec:obs}

We analyzed a flare observed simultaneously in white light, UV, and X-ray. Both in the WL and X-ray range the flare parameters were determined and compared with each other. X-ray observations also included spectroscopic observations.

\subsection{TESS data}

The white light data used in this work are from the Transiting Exoplanet Survey Satellite (TESS) \citep{2014SPIE.9143E..20R}. TESS mission started in 2018 and its main goal was to detect Earth-size planets orbiting nearby stars. The TESS bandpass is very wide and covers the range from 600 nm to 1000 nm.

\begin{figure}[H]

\centering
\includegraphics[width=\textwidth]{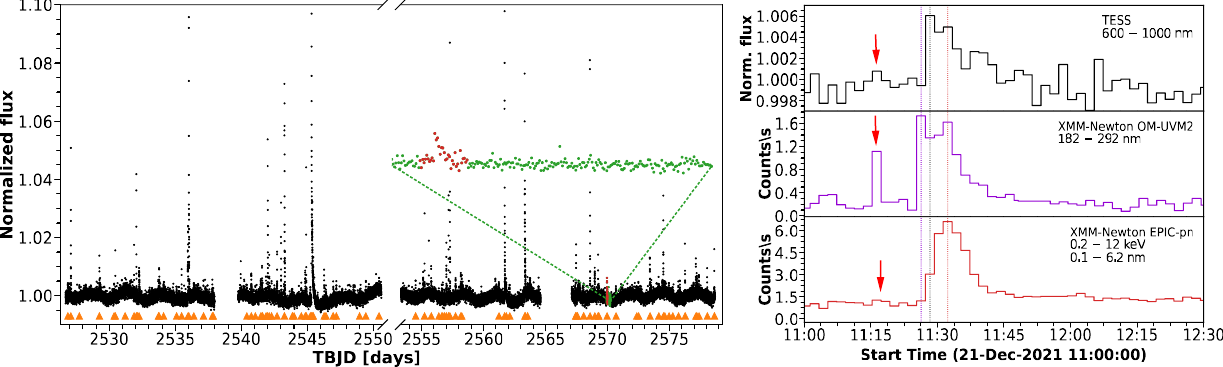}
\caption{\textit{Left panel:} The TESS light curve of Wolf 359 (sectors 45 and 46). Detected flares are marked with orange triangles, XMM-Newton observational interval in green, and the analyzed flare in red. \\ 
\textit{Right panel:} The light curve of the flare on Wolf 359 in white light, UV, and X-ray (0.2 - 12 keV). TESS two-minute cadence data are shown in black and XMM-Newton data binned to two minutes are in violet and red. The vertical colored lines mark the maximum of the flare in different energy ranges (black - white light, violet - UV, red - X-rays). The maximum of the flare in the white light range was at 11:28 UT, in the ultraviolet at 11:26 UT, and in X-rays at 11:33 UT. The pre-flare are marked as red arrows.
}
\label{fig:xmmtess}
\end{figure}

Wolf 359 was observed by TESS in November and December 2021 (sectors 45 and 46).  Figure \ref{fig:xmmtess} shows the changes in the star's brightness over time. The specific observations were obtained from the Mikulski Archive for Space Telescopes (MAST) at the Space Telescope Science Institute \citep{TESSFAST} \citep{TESSALL}.  It provides light curves with a time cadence of 20 seconds and two minutes. To analyze stellar flares we used Pre-search Data Conditioning SAP (PDCSAP) flux. We detected a total of 103 flares in both sectors of observations. On average, this star produces more than two flares per day.

We analyzed the flare observed on 21 December 2021. The flare is marked in red in Figure \ref{fig:xmmtess}. The right panel shows the light curves of the flare observed in white light (black), UV (violet), and X-ray (red) with a time resolution of two minutes. The event started at about 11:10 UT and ended at 12:15 UT. The vertical colored lines mark the maximum of the flare in different energy ranges.

\subsection{XMM-Newton data}

We used X-ray data from XMM-Newton \citep{2001A&A...365L...1J}. The observation data files were obtained from the XMM-Newton Science Archive\footnote{\url{https://nxsa.esac.esa.int/nxsa-web/##search}} (ObsID 0891802401). This space observatory has three simultaneously operating systems of instruments, which are European Photon Imaging Cameras (EPIC) \citep{2001A&A...365L..18S, 2001A&A...365L..27T}, The Reflection Grating Spectrometers (RGS) \citep{2001A&A...365L...7D}, and The Optical Monitor (OM) \citep{Mason_2001}. We used data from the pn-CCD camera. The time resolution of these observations is less than a second, but in our analysis, we used binned data due to poor signal statistics. Data from RGS were used to obtain the spectra of the flare. OM is a 30 cm Ritchey–Chrétien telescope. Its bandpass covers the range of optical and ultraviolet light (170 nm - 650 nm). The flare was observed in the fast mode with the UVM2 filter that covers the range from about 182 to 292 nm. We analyzed data from XMM-Newton using the Science Analysis Software (SAS version 20.0.0). 

\section{Analysis of the flare}  \label{sec:analysis}


Figure \ref{fig:xmmtess} (right panel) shows the differences in light curves of the flare in three wavelengths. The flare's maximum in the ultraviolet range was at 11:26 UT, in white light at 11:28 UT, and in X-rays at 11:33 UT. Before the analyzed flare, around 11:16 UT, a pre-flare can be observed marked as red arrows. At that time, the largest increase in the signal appeared in the white light and ultraviolet ranges, and also a weak structure was observed in X-rays. We are unable to determine if this preflare was physically related to the analyzed flare or not. The main event started suddenly after 11:25 UT in all ranges. The growth time in WL and UV was less than two minutes but in X-rays, it was four times longer. The decay times in each case were similar and were over half an hour. The decay time is defined as the time from the peak of the flare to the moment when the signal reaches the calculated background level. An interesting observed feature of the flare was that the X-ray (coronal) emission peaks about 6 minutes later than emission in UV and TESS bandpasses. This delay likely arose from the difference of characteristic reaction times in the stellar corona and chromosphere. Similar behavior is observed for the solar white-light flares, where the photospheric emission is well correlated with the emission of hard X-rays and proceeds the maximum in soft X-rays \citep{2015ApJ...802...19K,2016ApJ...816....6K}. Such a scheme is consistent with the model of a flare heated by the streams of non-thermal high-energy electrons \citep{2017ApJ...847...84F}. For the Sun the electrons with energies about 50 keV reach the chromosphere and the upper photosphere, where their energy is partly emitted in hard X-rays and partly deposited to the plasma and then radiated out in the white-light fast due to high plasma radiative losses. The low-energy part of non-thermal beam heats up the coronal part of the loop and radiates out in soft X-rays. The physical parameters of the coronal part change in the scale of thermodynamical time defined by \cite{1991A&A...241..197S} which is usually of a few to dozens of minutes.


The parameters of the whole event (pre-flare and flare) from the TESS light curve were obtained using our software WARPFINDER \citep{2022ApJ...935..143P}. The duration of the event was 64 minutes. The growth time was 24 minutes and the decay time was 40 minutes. Duration and decay times' values were typical of results obtained from a statistical analysis of more than 100,000 flares detected during the first three years of TESS observations \citep{2022ApJ...935..143P}. For example, the maximum of the distributions of all flare durations was approximately 50 minutes. However, due to the presence of a pre-flare, the software-estimated growth time of the analyzed event was longer than the typical value of about 10 minutes. The energy of the flare was estimated at $1.1 \pm 0.2 \times 10^{31}$ erg using the methods of \cite{2013ApJS..209....5S} and \cite{https://doi.org/10.1002/asna.200710756}. Compared to other stellar flares, this was very low energy. For such weak events, there are problems with automatic detection. In the case of Wolf 359 it was possible due to the star's low effective temperature and the small distance from the Sun. The median value of flares' amplitudes on this star is very close to the amplitude of the analyzed flare. The studied flare has durations and energies typical of most flares observed on this star. In all the TESS data, we detected flares with energies ranging from $10^{30}$ erg to almost $10^{33}$ erg. The longest and strongest flare lasted about four hours and had an energy of $7 \times 10^{32}$ erg.

\section{Results} \label{sec:res}

\subsection{Flare profiles}

\begin{figure}[H]
\gridline{\fig{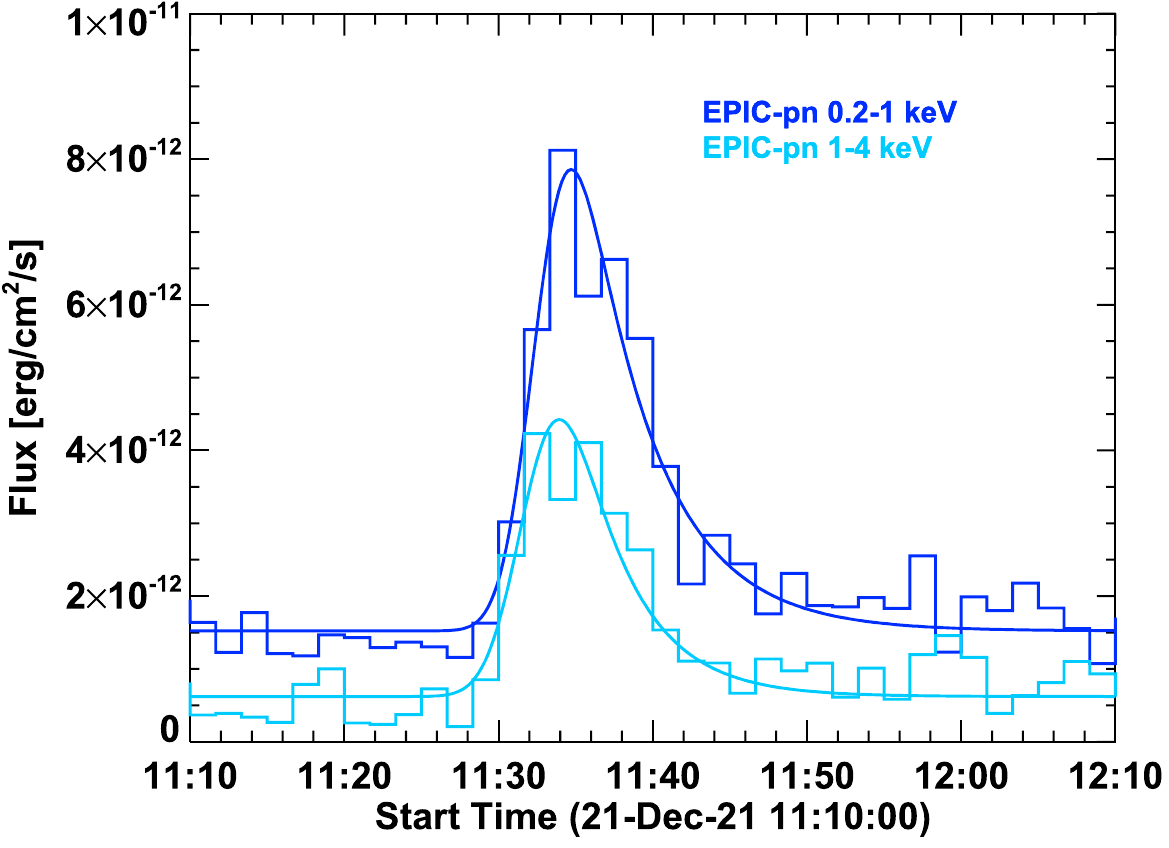}{0.506\textwidth}{}
          \fig{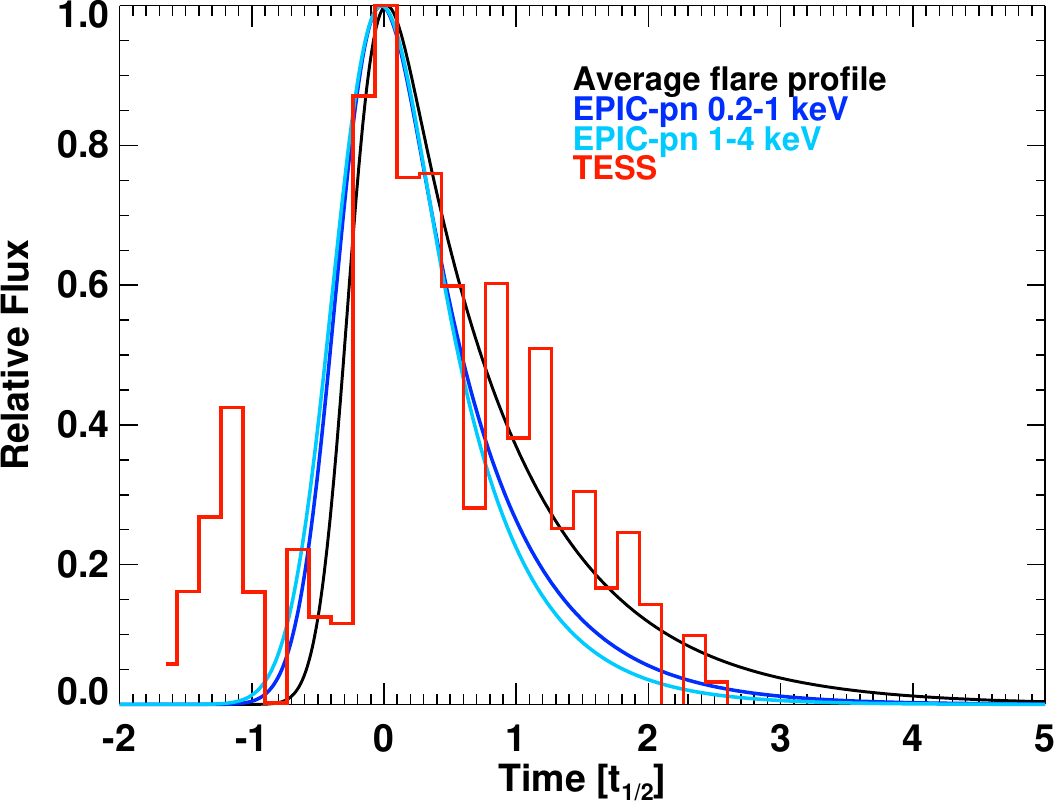}{0.48\textwidth}{}}
\caption{\textit{Left:} The light curves and flare profiles in two X-ray ranges. The range from 0.2-1 keV is shown in dark blue, and 1-4 keV in light blue. The time binning of the data was 100 second. \\ 
\textit{Right: }The flare EPIC-pn profiles (light and dark blue) and the TESS 100-second light curve (red) compared with the average flare single profile from \cite{2022ApJ...935..143P} (black).
}
\label{fig:profiles}
\end{figure}

Based on the X-ray data from EPIC-pn, we determined the fluxes during the flare in two ranges 0.2-1 keV and 1-4 keV. The time resolution of the  data was 100 seconds.The higher resolution lightcurve in 0.2-1 keV range, which show more details is available in Appendix \ref{sec:App2} (Fig. \ref{fig:diagnosticdiagram}). A time profile described in \citet{2017SoPh..292...77G} and  \citet{2022ApJ...935..143P} was fitted to each of them and to the TESS data. Figure \ref{fig:profiles} shows the EPIC-pn fluxes and the obtained flare profiles. The duration of the flare in the 0.2-1 keV range is 32 minutes and in the 1-4 keV range it is 26 minutes. These values are smaller than the duration of the flare observed in white light and ultraviolet. Growth times for both profiles are similar, about 8 minutes. The profile peak in the higher energy range is at 11:34 UT, about half a minute earlier than in the lower energy. This time is consistent with the maximum of the flare in the X-rays from Figure \ref{fig:xmmtess}. The decay time in the 0.2-1 keV range is longer than in the 1-4 keV and it is respectively 24 and 20 minutes.

The obtained EPIC-pn profiles and TESS 100-second light curve were compared with the average single flare profile form \citet{2022ApJ...935..143P}. Right panel of Figure \ref{fig:profiles} shows profiles scaled to  $t_{1/2}$,which is the full width of the light curve at half-maximum \citep{2013ApJS..207...15K}. There are differences between the X-ray (light and dark blue) and the white light profiles (red and black) at the decay phase ($t>t_{1/2}$). They may result from problems with determining the pre-flare background. The Wolf 359 flare's light curve (red) is consistent with the average single profile (black) obtained from observations of more than 100 000 flares on stars of various spectral types.

\subsection{Flare parameters}

From the XMM-Newton EPIC-pn X-ray data, we determined the temperature and emission measure during the flare. We used a single-temperature model and the filter-ratio method, i.e. we calculated the temperature from the ratio of fluxes in two spectral ranges. We decided on this simplest diagnostics due to the poor spectral statistics. 
We found that the steepest monotonic function is obtained for flux ratios in the ranges 0.2-1 keV and 1-4 keV. Above 4 keV the errors in the observed fluxes are too large to be useful. 
The theoretical dependence of fluxes on temperature was obtained using the SolarSoft (SSW) Chianti package in version 10.0.1 \citep{Dere_1997A&AS..125..149D,2021ApJ...909...38D} with solar photospheric elements' abundances and original Chianti ionization equilibrium.

\begin{figure}[H]
\gridline{\fig{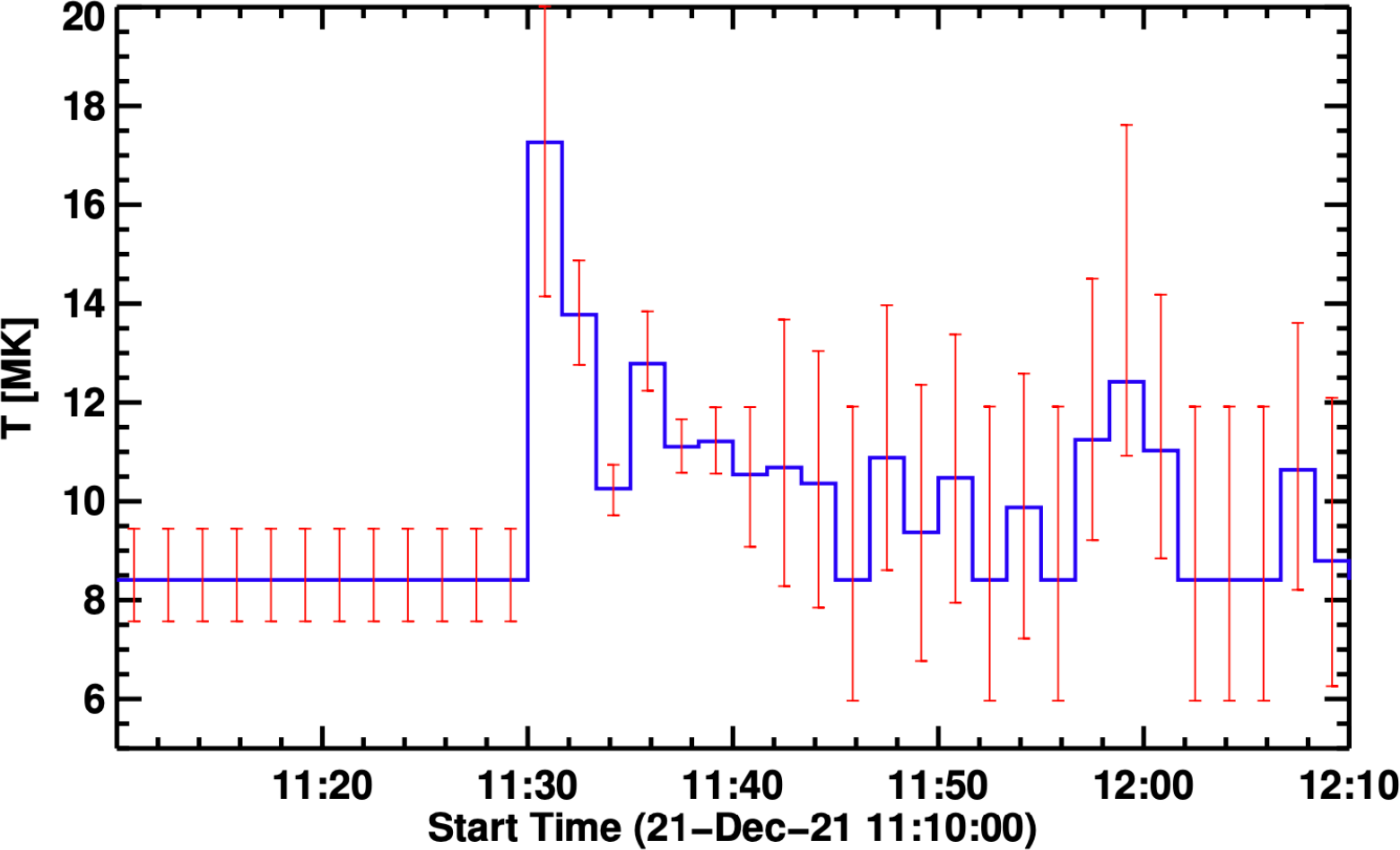}{0.49\textwidth}{}
          \fig{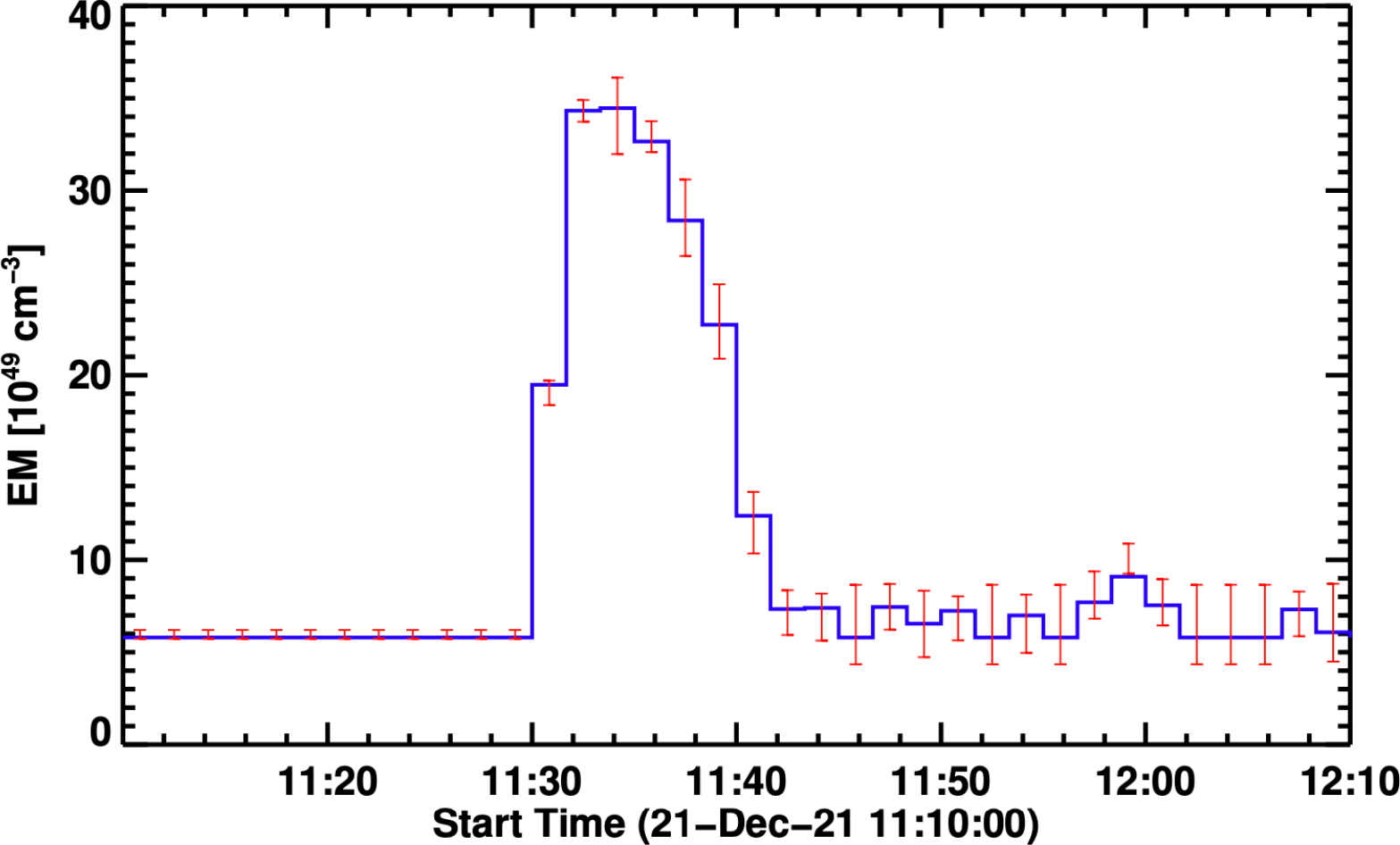}{0.49\textwidth}{}}
\caption{\textit{Left: } The temperature estimated during the analyzed flare from EPIC-pn X-ray data. The straight line before the beginning of the flare shows the mean temperature. The actual values could not be reliably determined due to the high data noise. \\ 
\textit{Right:} The emission measure estimated during the analyzed flare from EPIC-pn X-ray data. The straight line before the beginning of the flare shows the mean emission measure. The actual values could not be reliably determined due to the high data noise.
}
\label{fig:temperature_EM_MS}
\end{figure}

Figure \ref{fig:temperature_EM_MS} shows the changes in temperature  and emission measure during the flare. The maximum observed temperature is 17.25 MK. During the entire flare, the temperature exceeds 7 MK. Reliable temperature estimates are only possible for data from 11:26 UT to 11:44 UT. Figure \ref{fig:temperature_EM_MS} shows also an estimate of the emission measure during the flare. It is at least $1 \times 10^{50}$ cm$^{3}$, up to $3.5 \times 10^{50}$ cm$^{3}$ at the flare maximum.

\begin{figure}[H]
\includegraphics[width=1.0\textwidth]{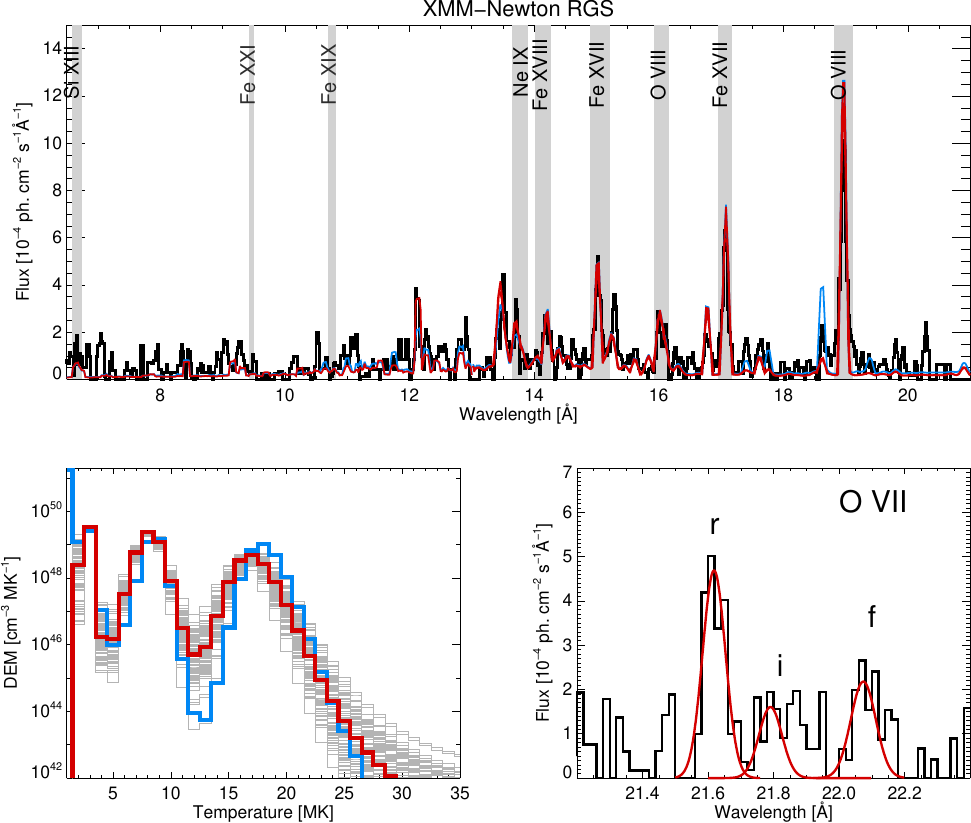}
\caption{\textit{Upper:} Average observed  spectrum  of the analyzed flare (in black). Gray stripes correspond to the spectral bands used for the DEM determination. The colors show the spectra calculated based on DEM distribution  (Figure \ref{fig:spectra}) using two abundances sets  \cite{Scott_a_2015A&A...573A..25S,Scott_b_2015A&A...573A..26S}, in blue and \cite{2007A&A...468..221F}, in red. \\ 
\textit{Bottom left:} DEM distributions determined based on unperturbed line fluxes (colors) and the 100 Monte-Carlo realizations of DEM calculations (gray) when elemental abundances by \cite{2007A&A...468..221F} were used. Red and blue colors represent the results obtained using two abundance sets  \cite{Scott_a_2015A&A...573A..25S,Scott_b_2015A&A...573A..26S} and \cite{2007A&A...468..221F} respectively.\\
\textit{Bottom right:} The O {\sc{vii}} triplet in the RGS spectrum on 21 December 2021. Red lines are the Gaussian profiles fitted for each line.
}
\label{fig:spectra}
\end{figure}

The RGS spectra let us calculate the differential emission measure (DEM) distributions. We determined the shape of DEM using the Withbroe-Sylwester \citep{1980SoPh...67..285S} method that is  the  improved Maximum Likelihood iterative algorithm. The study is based on the absolute flux values containing line and continuum emission observed in  nine spectral bands without background subtraction. Figure \ref{fig:spectra} (upper panel) presents a part of the RGS spectrum (6--21~\AA) with spectral ranges used for a DEM analysis. The emission functions corresponding to the selected spectral ranges were calculated using Chianti 10.0.1 atomic code  for two sets of elemental abundances: \cite{Scott_a_2015A&A...573A..25S,Scott_b_2015A&A...573A..26S}  and  \cite{2007A&A...468..221F}.  The analysis was conducted in the temperature range from 1 to 40 MK with a temperature interval of 1 MK. Figure \ref{fig:spectra} (bottom left panel) 
shows also the obtained results based on the average spectra of the December 21, 2021 flare (11:25 UT - 13:20 UT). The blue and red colors represent the results obtained for different abundance sets, by \cite{Scott_a_2015A&A...573A..25S, Scott_b_2015A&A...573A..26S} and \cite{2007A&A...468..221F} respectively. 
The obtained DEM distributions are three components at temperature values of  3 MK, 7 MK, and 16--17 MK. Similar results were reported by \cite{2007A&A...468..221F} based on a study of data collected on May 2004 and December 2005. The temperature corresponding to the hottest DEM component agrees with parameters obtained using the isothermal model.
Poor count statistics give large values of ''formal'' uncertainties of flux values that strongly affect the uncertainties of DEM distribution. Therefore, we assumed 10\% flux errors. The gray thin lines show DEM distributions obtained for 100 Monte-Carlo realizations.

The analysis of the ratio between forbidden ($f$) and intercombination ($f$) lines in helium-like triplets is a valuable method for determining the electron density ($n_e$) of the plasma. This theory was originally proposed by \cite{Gabriel_1969MNRAS.145..241G} and is extensively used in literature e.g. \cite{Mewe_1978A&A....65...99M}, \cite{Pradhan_1981ApJ...249..821P}, \cite{Ness_2001A&A...367..282N}, \cite{2007A&A...468..221F}, \cite{Stelzer_2022A&A...667L...9S}.

The sensitivity of this ratio to electron density ($n_e$) varies depending on the specific ion and line triplet under consideration. When examining helium-like triplets formed at higher temperatures, there is a higher sensitivity to electron densities compared to triplets formed at lower temperatures. \cite{2009A&ARv..17..309G} reported specific density regimes for different ions, where the sensitive range for oxygen is $log(n_e) = 9.5-11.5$, for neon $log(n_e) = 11.0-13.0$, for Mg XI  $log(n_e) = 12.0-14.0$, and for Si XIII $log(n_e) =  13.0-15.0$.

Within its wavelength range (6-35 \AA) the RGS contains the Helium-like line “triplets” from N VI to Si XIII. However, it is worth noting that an in-depth analysis of the N VI, Ne IX, Si XII, and Mg XI triplet lines was not conducted due to the insufficient statistical quality. Consequently, the determination of density relies on the ratio of the O VII lines, which is the most prominent He-like triplet.

We fitted Gaussian profiles for each line (Figure \ref{fig:spectra}, bottom right panel) 
and used the relation $f/i=R_0(1+n_e/N_c)$.  We assumed $N_c$~=~3.1$\times$10$^{10}$~cm$^{-3}$ and $R_0$~=~3.95 according to  \cite{Pradhan_1981ApJ...249..821P}.
The measured obtained value of $f/i$ ratio is 1.47 which gives $n_e$~$\sim$~5$\times$10$^{10}$~cm$^{-3}$. The density calculation was based on average spectra obtained during the whole observations available for 21 December 2021 collected during quiet periods and stellar flares.

\section{Discussions} \label{sec:discussions}

We analyzed observations of a flare on the star Wolf 359 simultaneously in white light, UV, and X-ray. The study was possible thanks to the unprecedented sensitivity of the instruments located on the XMM-Newton satellite (UV and X-ray range) and TESS (WL range).  For the first time a stellar flare with energy comparable to an X-class solar flare is analyzed on this star. 
Observations of the solar-like flare on Wolf 359 are a very important element on the basis of which the parameters and energy of the flare on another star with energies comparable to medium-strong solar flares were determined. By observing and analyzing such events we can test a hypothesis that we are dealing with similar versions of the same physical phenomena.

The obtained EPIC-pn profiles and TESS 100-second light curve were compared with the average single flare profile from \citet{2022ApJ...935..143P}. The Wolf 359 flare’s light curve (red) is consistent with the average single
profile (black) obtained from observations of more than 100 000 flares on stars of various spectral types (see Figure \ref{fig:profiles}). The observed differences between the X-ray (light and dark blue) and white light profiles (red and black) at the decay phase ($t > t_{1/2}$) may result from problems with determining the pre-flare background.

From the XMM-Newton EPIC-pn X-ray data, we determined the temperature and emission measure (a single temperature model) during the flare. The maximum observed temperature is 17.25 MK. Estimates of the emission measure during the flare are at least $1 \times 10^{50}$ cm$^{-3}$, up to $3.5 \times 10^{50}$ cm$^{-3}$ at the event maximum. The RGS spectra allowed us to calculate the differential emission measure distributions (DEM). The obtained DEM distributions are three components at temperature values of  3 MK, 7 MK, and 16 MK--17 MK (see Figure \ref{fig:spectra}). The  temperature calculated using the isothermal model agrees well with the temperature corresponding to the hottest component of DEM. The other two components can be interpreted as the temperature of the active regions (about 7 MK) on the surface of the star and the temperature of about 3 MK representing the quiescent corona. Total emission measure derived from DEM for abundance set from paper \cite{Scott_a_2015A&A...573A..25S,Scott_b_2015A&A...573A..26S} has a value $\rm 9.2\times10^{49}\, cm^{-3}$. This value is lower compared to the determination from the filter-ratio method due to the spectrum accumulation time covering the entire duration of the flare.

The analysis of  helium-like  triplet O {\sc{vii} allows us to determine the  plasma electron density ($n_e$~$\sim$~5$\times$10$^{10}$\,cm$^{-3}$) (see Figure \ref{fig:spectra}). This value is about one order lower than the estimated average density in the flare loop ($N_e=7.83\times10^{11}\,$cm$^{-3}$, Appendix \ref{sec:App2}}). The reason for obtaining a lower density from the ratio of forbidden and intercombination lines was due to the accumulation time of the spectra. These spectra covered the available observations for December 21, 2021, with an interval of approximately 6 hours. The observations were collected during periods of quiescence and stellar flares. Additionally, it should be noted that O VII is a relatively low-temperature ion (see \cite{2009A&ARv..17..309G}).

An attempt was made to determine the flare class by various direct and indirect methods (more details in Appendix \ref{sec:App1}). The GOES class calculated from the 1-4 keV flux for the EPIC instrument (which is the closest to the GOES 1-8 \AA~measurement range) on board the XMM-Newton  was X4.3.
We also used the single-temperature model and the filter-ratio method to calculate the temperature and emission measure from the ratio of fluxes in two spectral ranges 0.2-1 keV and 1-4 keV. Having derived the maximum temperature and for the same time emission measure, we could calculate the GOES flux in the range of 1-8 \AA. The resulting flare class in this case was X5.7. All other methods are indirect and after applying them we get a lower flare class, based on the energy radiated in the rise and decay phase (X3.7), based on the DEM (X2.1), and using the bolometric energy estimated from TESS with a large error - $\rm M2.1_{B9}^{X31}$.

Using X-ray data, the parameters of the flare loop were determined from the rise and decay phases of the light curve of the phenomenon. Then we compared the determined geometric parameters of the loop, such as half-length, cross-section area, and volume of the flare loop on the Wolf 359 star to typical sizes/parameters for solar flares. For this purpose, the data from \cite{Warmuth_2013} were used, in which the dependence of the geometric parameters of the flare loops on the GOES class was analyzed.

We compared two methods of estimating the flare size and physical plasma parameters, one based on the analysis of the decay phase developed by  \cite{1997A&A...325..782R} and the second based on the analysis of the rise phase shown by \cite{2001AdSpR..26.1785P}. Results of both methods are in good agreement in the range of $L_0$ from $1.41\times 10^9$ cm to $1.55\times 10^9$ cm. This allowed us to determine the physical parameters of the flaring plasma (see Appendix \ref{sec:App2}) which appear to be very similar to the same parameters in strong solar flares.

In each of the three analyzed parameters, half-length of the loop, volume, and cross-section area for the analyzed flare on the Wolf 359 star, taking into account errors, fit well with the observed relationship for solar flares. The obtained data resulting from the comparison of values indicate good agreement with the determined GOES class for the analyzed flare which is roughly X3.7 - X5.7 (see Fig.\ref{fig:l0}).

\begin{figure}[H]
    \centering
    \includegraphics[width=0.64\textwidth]{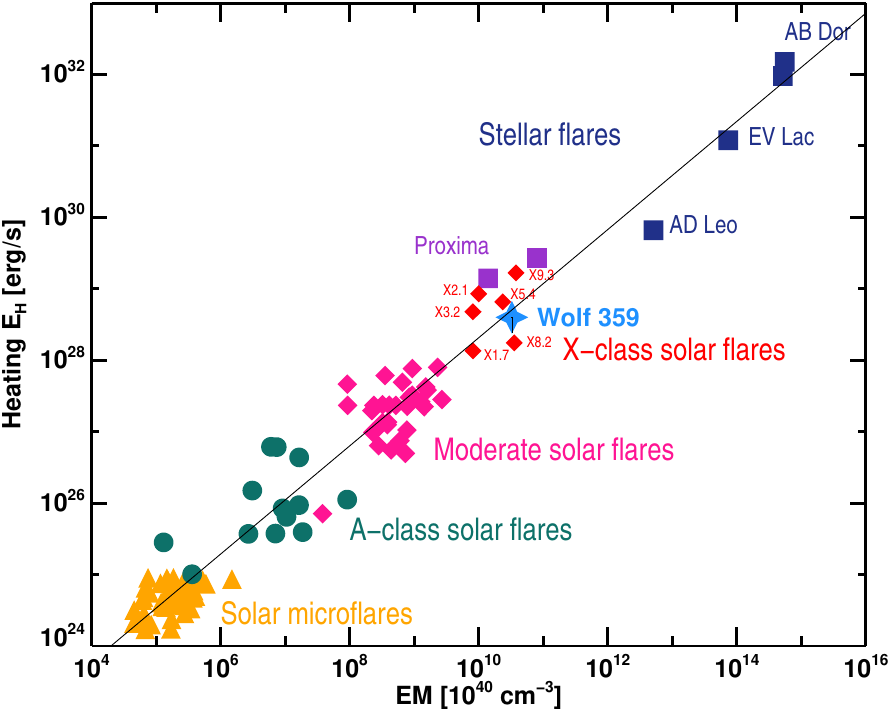}
    \caption{The location of the analyzed flare on the diagram $EM$ vs heating ($E_H$). The added X-class solar flares are marked red. Flare on Wolf 359 is marked with a blue star. The solid line shows the new $EM - E_H$ relationship based on all flares.}
    \label{fig:Em_EH}
\end{figure}

Figure \ref{fig:Em_EH} presents the 
 relation EM - $\rm E_H$ for events from \cite{Pres_2005} extended with X-class solar flares.
The added X-class flares have been plotted and described with the appropriate flare class.
The position of the analyzed stellar flare on the Wolf 359 star is the closest to the class X5.4 solar flare, which agrees well with the class estimate from the single-temperature method (X5.7) or the direct X-ray flux estimate from the EPIC instrument in the 1-4 keV range (X4.3). 
In addition, the presented EM - $\rm E_H$  relationship was improved using observations of X-class solar flares and the analyzed flare on Wolf 359. The equation of the new line is $\log(E_H) = -9.528 + 0.757 \log(EM)$ and has been drawn in the figure with a solid line. The error of the slope value is 0.0147, which means that it has changed by only 1$\sigma$ in relation to the previous determination. For solar and stellar flares, the determined relation works very well for heating energy rate ($\rm E_H$) in the range of $10^{24}-10^{33}\,\rm erg \; s^{-1}$ and for emission measures in the range of $10^{44}$ to $10^{56}\,\rm cm^{-3}$. This means the relationship between the heating energy rate and the emission measure is preserved for both solar and stellar flares. This implies a legitimate conclusion made at the beginning of the work that the processes taking place in stellar flares are consistent with a solar-type flare mechanism. A good example is the analyzed event, which has a flare class similar to strong solar flares. The determined geometrical parameters of the phenomenon do not differ from the values of analogs occurring on the Sun. The determined minimum magnetic field estimate in the flare loop also has a value typical for solar flare loops. The only difference is the much higher average density in the flare loop, which is about one order higher than for typical solar phenomena. This may be due to the fact that the star is a red dwarf with a larger log({\it g}) and a smaller pressure scale height.

\begin{acknowledgments}

This work was partially supported by the program "Excellence Initiative - Research University" for years 2020-2026 for University of Wrocław, project no. BPIDUB.4610.96.2021.KG.

The authors acknowledge the XMM-Newton and the TESS consortium for providing excellent observational data.

\end{acknowledgments}

\appendix

\section{What is the X-ray solar-like class of the flare?} \label{sec:App1}

The radiated energy in the rise phase ($E_{r}=7.20\times10^{29}\,$erg) and the flare decay phase ($E_{d}=4.65\times10^{29}\,$erg) was estimated separately using methods described in Appendix \ref{sec:App2}. The total radiated energy is $E_{tot}=1.18\times10^{30}\,$erg. We can use formula (4) from \cite{Cliver_2022} to go from soft X-ray radiated energy to peak brightness. The flare class we obtained was X3.7. The GOES class calculated directly from the 1-4 keV flux for the XMM-Newton EPIC-pn (see Fig.\ref{fig:profiles}), which is closest to the GOES 1-8 \AA \ measurement range, was class X4.3. We used routine \emph{goes\_fluxes.pro} in the SolarSoftWare (SSW) written in Interactive Data Language (IDL) for converting emission measures and temperature into GOES 1-8 \AA \: flux. For the DEM distribution, the class of the flare was about X1.0. This determination was obtained for emission lines and continuum from the RGS spectrum. It may be underestimated due to the averaging used spectrum over the duration of the flare. After taking into account the flux-averaging effect, the resulting class is X2.1. We used a single-temperature model and the filter-ratio method to calculate the temperature and emission measure from the fluxes in two spectral ranges 0.2-1 keV and 1-4 keV. Having derived the maximum temperature $T=17.26\,$MK at 11:30:50.0 UT and for the same time emission measure $EM=1.95\times10^{50}\,$cm$^{-3}$ we could use again routine \emph{goes\_fluxes.pro} to calculate the GOES flux in the range of 1-8 \AA. The resulting flare class was X5.7.

The energy of the flare was also estimated from TESS data using two methods. The bolometric energy was about $1.1 \times10^{31}\,$erg. We used the \cite{Cliver_2022} relation (3) to convert bolometric energy to energy emitted in X-ray. The estimated value was $5.1\times10^{28}\,$erg and taking into account the errors, the minimum energy value was $4.53 \times10^{26}\,$erg and the maximum was $1.26 \times10^{31}\,$erg. By recalculating the energy ranges obtained
from soft X-ray radiated energy to peak brightness we can use again formula (4) from \cite{Cliver_2022}. The resulting range of GOES classes is large, from B2.9 for the lowest energy, through M2.1 for the middle value, to X31 for the maximum value. It is visible that the bolometric energy estimates from the TESS data seem to be underestimated or the relation on the basis of which the GOES class was determined is not entirely correct. It should be emphasized that the authors tested it not only on solar but also on stellar data. There is also the third explanation that the relation is inaccurate, hence such a large scatter in data.

To check our calculations, we decided to supplement the relation EM - heating for events from \cite{Pres_2005} of X-class solar flares (see Fig.\ref{fig:Em_EH}). We used the catalog of flares of the XRT instrument (\cite{Golub_2007}) on the Hinode satellite (\emph{https://hinode.isee.nagoya-u.ac.jp/flare\_catalogue/}) and selected X-class flares. We took 5 of 61 phenomena for further analysis. The availability of unsaturated images in the Be\_thick or Be\_med filter was the criterion. An additional factor was the simple structure of the flare that allowed us to determine the volume and loop length of the event. GOES data were used to determine the maximum temperature and at the same time the value of the emission measure for all flares. It was needed to calculate the heating value. The obtained results were plotted in Figure \ref{fig:Em_EH}, where each point was additionally described with the appropriate flare class.
The position of the analyzed stellar flare on the Wolf 359 star is the closest to the class X5.4 solar flare, which agrees well with the class estimate from the single-temperature method (X5.7) or the direct X-ray flux estimate from the EPIC instrument in the 1-4 keV range (X4.3).
All other methods are indirect and after applying them we get a lower flare class, based on the energy radiated in the rise and decay phase (X3.7), based on the DEM (X2.1), and from the bolometric energy estimated from TESS with a large error from B9 to X31 (M2.1).


\section{Estimation and the uncertainties of flare loop parameters} \label{sec:App2}

\begin{figure}[H]
\fig{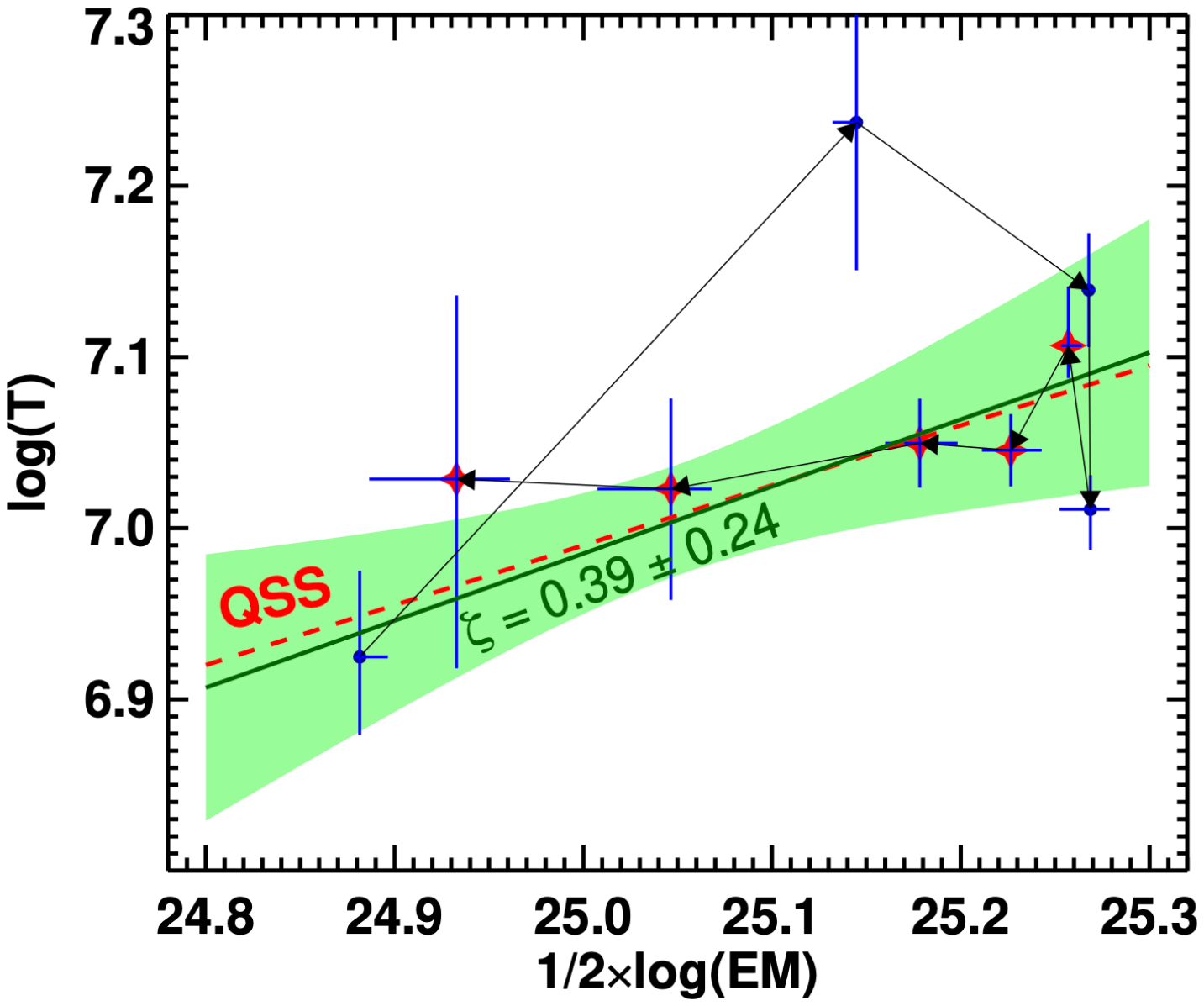}{0.8\textwidth}{}
\caption{The diagnostic diagram of the flare ($\log \sqrt{EM}$ vs. $\log T_{obs}$). The emission measure and temperature are expressed in cm$^{-3}$ and K respectively. Quantities in logarithms are dimensionless. The green line shows the best fit with the slope $\zeta = 0.39 \pm 0.24$ for the decay phase (red stars) and the red line represents the slope of the quasi-stationary decay ($\zeta_{QSS} = 0.35$). Green shaded region is the confidence interval. \\ 
}
\label{fig:diagnosticdiagram}
\end{figure}

The method from \cite{1997A&A...325..782R} allows us to estimate the loop semi-length along which the flare loses its thermal energy through conduction. The light-curve e-folding decay time is often affected by the sustained decay of heating. It can be determined from the slope $\zeta$ of the decay phase on the density-temperature diagram. The slower decay of heating, the closer the observed slope $\zeta$ to the quasi-stationary decay (QSS). For XMM/EPIC \cite{1997A&A...325..782R} estimates the QSS slope as $\zeta_a = 0.35$. In our case, the slope $\zeta$ is estimated as $0.39 \pm 0.28$ (see Figure \ref{fig:diagnosticdiagram}). This allows us to assume that the decay of the event was substantially affected by the slow deterioration of residual heating and the evolution was close to QSS. We take into account only the upper limit of $\zeta < 0.67$.  This allows us to estimate that the light-curve e-folding decay time, $\tau_d = 284 \pm 25\,$s, is longer than the thermodynamical time by the factor at least $F(\zeta) > 2.95$. This gives us the upper limit for $\tau_{th} < 104.7 \, $s, and the loop semi-length $L_0 < 1.55\times 10^9\,$cm, what is $L_0 < 0.16\, R_*$. 

\begin{figure}[H]
\gridline{\fig{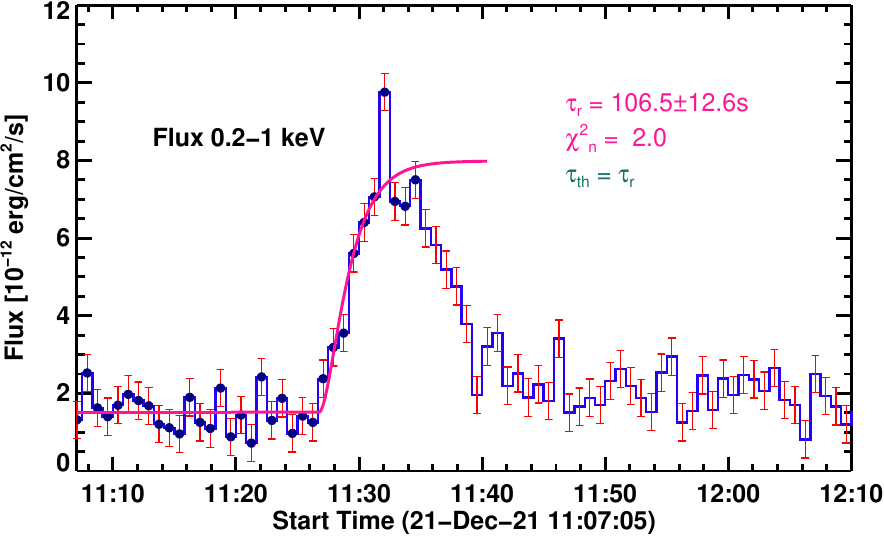}{0.49\textwidth}{}
           \fig{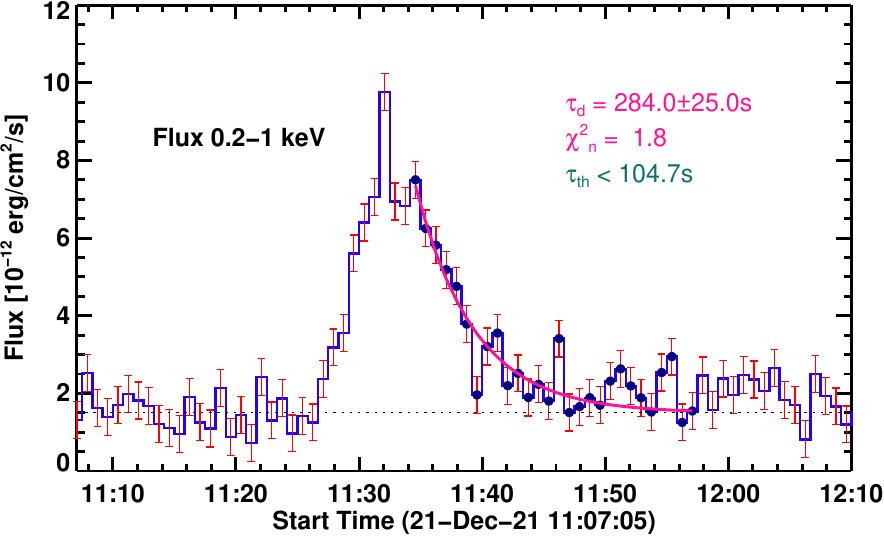}{0.49\textwidth}{}}
\caption{\textit{Left pannel:} The best fit to the light curve of the rise phase. The data are binned every 50 sec. The estimated thermodynamical time is $\tau_{th} = 106.6 \pm 12.6\,$s and the saturation flux $F_{ss} = 6.47 \pm 0.35\,$ cts s$^{-1}$. The black dots indicate the points used in the fit function for the rise phase.\\
\textit{Right pannel:} The best fit of the exponential decay to the light curve of the decay phase. The data are binned every 50 sec. The estimated thermodynamical time should not be greater than $\tau_{th} = 104.7\,$s. The black dots indicate the points used in the fit function for the decay phase.
}
\label{fig:Reale_Pres}
\end{figure}

\begin{figure}[H]
    \centering
    \includegraphics[width=0.6\textwidth]{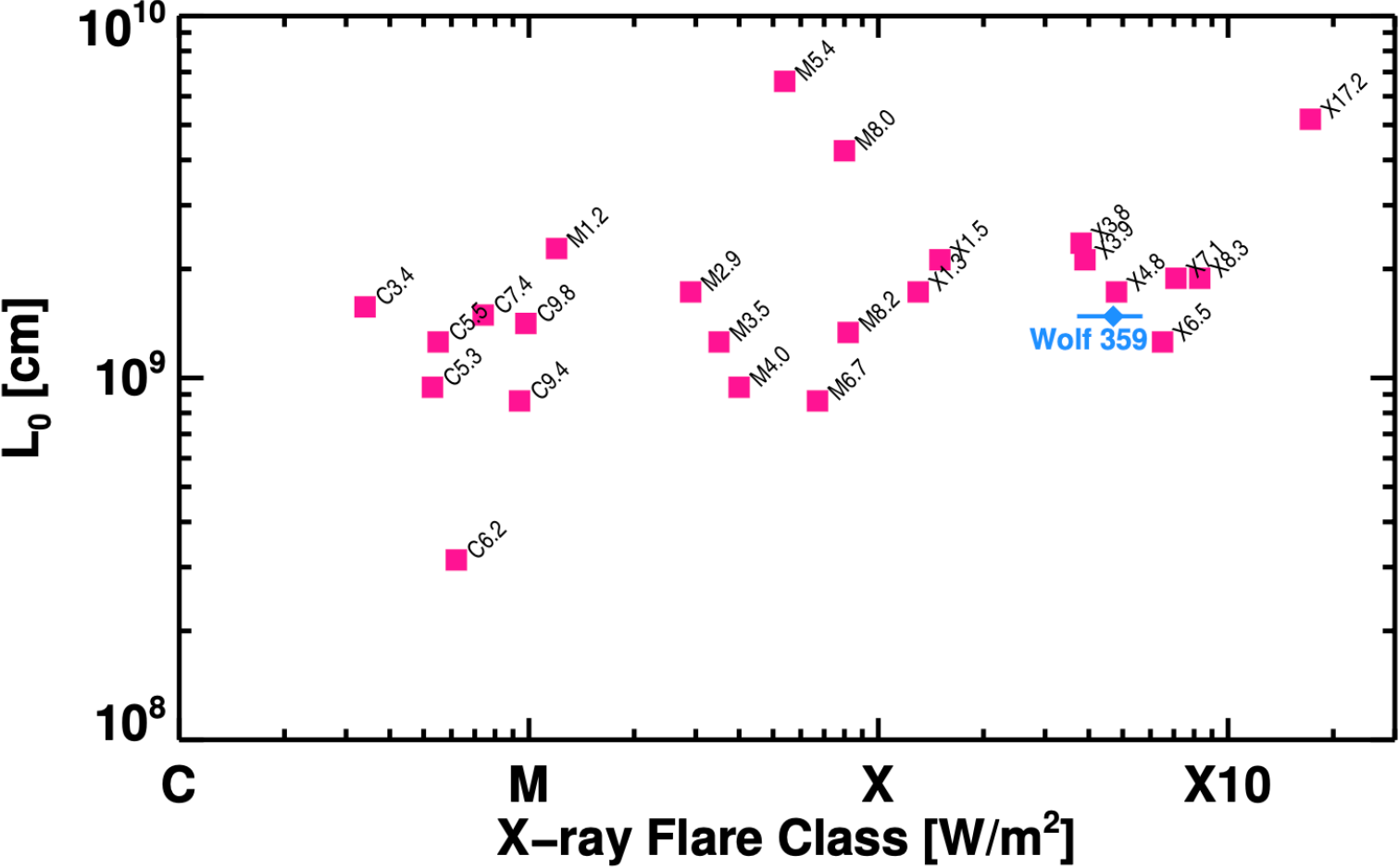}
    \caption{Relationship between the GOES flare class and the half-length of the flare loops for 24 solar events analyzed in \cite{Warmuth_2013} represented by red squares. At each point, there is also a description of the GOES flare class. The error ranges are marked in blue and the $L_0$ value for the analyzed flare on the star Wolf 359 is marked with the diamond symbol.}
    \label{fig:l0}
\end{figure}

\cite{1993A&A...267..586S} showed that the evolution of thermodynamical measure ($\eta=T\sqrt{EM}$) on the flare rise phase may be described by $\eta = \eta_{ss}(1-exp(-t/\tau))$, where $\tau$ is a close approximation of the thermodynamical time $\tau_{th}$ \citep{1991A&A...241..197S}. \cite{2001AdSpR..26.1785P} showed the method to estimate $\tau_{th}$ from the analysis of the light curve on the flare rise phase. We achieve the best fit with the parameters $\tau_{th} = 106.6 \pm 12.6\,$s (see Fig.\ref{fig:Reale_Pres}). 
This value transforms to the loop semi-length estimation $L_0=(1.59 \pm 0.18)\,\times 10^9\,$cm. Results of both methods are in good agreement in the range of $L_0$ from $1.41\times 10^9$ cm to $1.55\times 10^9\,$cm. The half-length of the loop estimates by two independent methods indicate quite solar values. In Figure  \ref{fig:l0} where we present the dependence of the flare class and the half-length of the flare loops occurring in the Sun.

It is possible to estimate the average density in the loop at the flare peak having determined the half length of the loop and the maximum temperature ($T_e = 17.25 \, $MK). During the rise phase the flaring loop passes for the pre-flare equilibrium to the new one, with the higher heating and the temperature. This allows us to estimate the loop density at the flare maximum from the  \cite{1978ApJ...220..643R} scaling law for the coronal loop at the new equilibrium with the higher, flare heating. The calculated value was $N_{e}=7.83\times 10^{11}\,$cm$^{-3}$. We estimated the electron pressure $ P_e=6970.5$ dyn cm$^{-2}$. Using the balance between the electron pressure $P_e$ and the magnetic pressure ($ P_m$), it is possible to calculate a lower estimate of the magnetic field ($P_m\geq 418.6\,$Gs). The volume of the flare loop ($V$) was calculated using electron density and emission measure ($EM=3.26\times 10^{50}\rm\, cm^{-3}$), it was $V=5.32 \times 10^{26} \rm \, cm^{3}$. The cross-section area ($ A_{cs}$) of the loop was determined using the volume and half length of the flare loop. Its value was $ A_{cs}=1.67\times 10^{17} \rm \,cm^2$. Using the values of the maximum temperature, half loop length, and volume determined above, the heating energy rate $E_H=3.99 \times 10^{28}\rm\, erg\,s^{-1}$ was estimated.

An important issue in the analysis of this flare was the high uncertainties of the temperature estimate during the rise phase, covering the range from 14 MK up to 80 MK. It was caused by a low signal-to-noise ratio in the observational data. These uncertainties propagate to estimates of sizes and physical parameters of flare plasma. This dependence is weak in the case of loop length but gives substantial changes in other estimations. The least sensitive to this problem is the value of the total heating rate in the whole flare loop. The uncertainties become relatively small when we search for an agreement between two methods of estimating the loop size. The loop semi-length ($L_0$) vary from $1.41\times 10^9$ to $1.55\times 10^9$ cm. The plasma density ($N_{e}$) at the flare maximum varies from $5.4\times 10^{11}$ to $7.3\times 10^{11}\ \rm{cm^{-3}}$ and the plasma pressure from $3.8\times 10^3$ to $6.1\times 10^3\ \rm{dyn \, cm^{-2}}$. The estimated flare volume ($V$) varies from $6.2\times 10^{26}$ to $7.8\times 10^{26}\ \rm{cm^3}$ and the total heating rate $E_H$ varies in the range $2.4-4.0\times 10^{28} \, \rm erg \,s^{-1}$.

We compared the geometric parameters of the flare loop on the Wolf 359 to typical sizes for solar flares. We used data from \cite{Warmuth_2013}. Figure \ref{fig:l0} shows the values of half-length of the loop ($L_0$) for 24 solar flares in relation to the GOES class. The location of the determined half-length of the loop on the Wolf 359 star fits well with the observed relationship for solar flares. For the estimated $\rm L_0$ value, the corresponding classes of solar flares are in the range X4.8 to X5.5. In the case of the other relationships volume - flare class and cross-section area - flare class, the values corresponding to the values determined for the flare on the Wolf 359 star are located in the vicinity of solar flares with the class about X5.0. The obtained result from the comparison of values indicates good agreement with the determined GOES class for the analyzed flare roughly which is X3.7 - X5.7.


\bibliography{sample631}{}
\bibliographystyle{aasjournal}



\end{document}